# Bulk Material Based Selective Infrared Emitter for Sub-Ambient Daytime Radiative Cooling

Yue Yang,[1,#,*] Linshuang Long,[2,#] Sheng Meng,[1] Nicholas Denisuk,[2] Liping Wang,[2,*] and Yonggang Zhu[1,3]

[1] School of Mechanical Engineering and Automation,
Harbin Institute of Technology, Shenzhen, 518055, PR China

[2] School for Engineering of Matter, Transport & Energy,
Arizona State University, Tempe, AZ 85287, USA

[3] School of Science, RMIT University, Melbourne, VIC 3001, Australia

[#] These authors contributed equally to this work.

[*]Corresponding author. Email: yangyue2017@hit.edu.cn (Y.Y.); liping.wang@asu.edu (L.W.)

Through passively emitting excess heat to the outer space, radiative cooling has been demonstrated as an efficient way for energy saving applications. Selective surface with unity emittance only within the atmospheric window as well as zero absorption within the solar spectrum is sought to achieve the best sub-ambient radiative cooling performance during the daytime. In this work, we proposed a bulk radiative cooler consisting of a 1-mm-thick lithium fluoride crystal coated with silver backing, which exhibits a solar absorptance of 4.7% and nearly ideal infrared selectivity with high emission exactly within the atmospheric transmission band (i.e., 8-13 μm). Excellent daytime cooling performance was demonstrated in an outdoor test with stagnation temperature below the ambient temperature by 5 °C under solar irradiance above 900 W/m$^2$ and a net cooling power of about 60 W/m$^2$ when the cooler is in thermal equilibrium with the ambient. As a bulk material with the highest ultraviolet-visible-near infrared transmittance, lithium fluoride crystal has been widely employed as the optical windows and mirrors in various applications. The proposed simple selective infrared emitter based on lithium fluoride would open up an innovative way to radiatively cool optical systems.

As a passive cooling method without energy input, the concept of radiative cooling is to radiatively dissipate the excess heat into the cold outer space through the atmospheric window.[1] It has great potential in applications like energy-efficient building, photovoltaic cooling, food refrigeration, and personal thermal management.[2-4] Historically, the idea of radiative cooling could be dated back to the 1950s.[5] Selective thermal emission within the infrared atmospheric window (8-13 μm) is required for efficient radiative cooling, while a low solar absorptance is needed to minimize solar heating during the daytime. Due to the supreme optical property of selective emission in the infrared region, the bulk materials of plastic,[6,7] oxide,[8,9] nitride,[10,11] and composite materials of pigmented paints [12,13] as well as gaseous materials [14,15] were the main objectives searched to achieve radiative cooling from the 1960s to 1990s. The coolers were usually composed of a layer of the bulk material with alumina/silver (Al/Ag) coated as the back reflector. However, because of the non-ideal selectivity in the infrared region and relatively high solar absorptance of these bulk materials, sub-ambient cooling effects were only achievable during the nighttime but not daytime.

It was until the year of 2013 when Rephaeli et al.[16] first theoretically demonstrated sub-ambient daytime cooling performance, through a calculation of radiative energy exchange between the environment and the cooler, a metal-dielectric photonic structure. The same group[17] also experimentally demonstrated sub-ambient cooling effect under direct sunlight with a multilayer based radiator. These pioneer works opened a field of designing micro/nanosized photonic structures to achieve sub-ambient daytime cooling, including multilayer films[18-20] and patterned surfaces.[21-23] However, the complexity and high cost in fabricating photonic structures significantly limit their wide applications. Later, radiative emitters with a simple fabrication process and the possibility of scalable manufacture, including a TPX layer embedded with glass nanoparticles,[24] combinations of a fused silica layer and a PDMS film,[20] and complete delignification and densification of wood,[25] were proposed for demonstration of daytime radiative cooling effect. All of these emitters occupy broadband high infrared emission and low solar absorptance in the meanwhile.

The reported best daytime cooling performance up to date from outdoor tests conducted under direct sunlight[20,24-28] is with a range of 4~8 °C below the ambient temperature and 40~130 W/m$^2$ net cooling power, considering the variations of environmental conditions and experimental setups. Although we cannot see much difference between selective and broadband emitters in the daytime cooling performance from the conducted outdoor tests, the selective emitters hold great advantages to achieve a much lower stagnation temperature, especially when the convective and conductive heat losses are eliminated.[2] Potentially, the selective emitter could obtain an ultra-low stagnation temperature of more than 85 °C below the ambient, while that for the broadband one is limited to 40 °C.[4] Experimentally, through blocking the direct sunlight and eliminating the convection loss in vacuum, Chen et al.[19] employed a selective emitter to gain a temperature drop up to 42 °C below the ambient, which was already beyond the limit of a broadband emitter even with ideal spectral emission. While photonic structure-based emitters possess advantages in achieving selective emission compared to the broadband emitter, they also face the challenge of the complicated fabrication process and high manufacturing cost.

In this work, we would like to introduce a selective radiator consisting of a single 1-mm-thick bulk lithium fluoride (LiF) crystal coated with Ag backing for efficient sub-ambient daytime cooling. Besides simple fabrication process, the proposed bulk LiF radiator occupies nearly ideal infrared selectivity with high emission exactly within the atmospheric transmission band (i.e., 8-13 μm), suggesting great potential to achieve a much lower stagnation temperature with well-controlled experimental conditions.[19] It is worth to note that Berdahl et al.[29] looked into the performance of bulk MgO radiator in the nighttime cooling application, and mentioned that LiF possessed the similar spectral selectivity. However, no study until the present work demonstrates a sub-ambient daytime cooling effect under direct sunlight with a bulk material made of LiF.

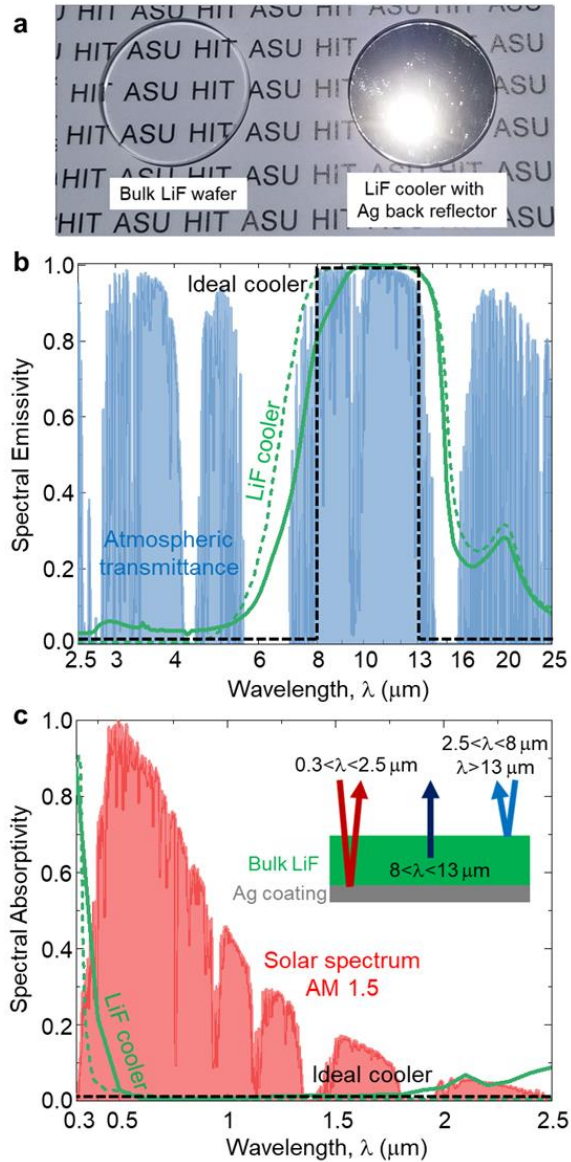

**Figure 1. LiF-based radiative cooler.** (a) Photos of the bare bulk LiF crystal (left) and the LiF cooler with Ag back reflector (right). Bulk LiF crystal is highly transparent for the visible wavelength range, while the LiF cooler, made of bulk LiF and Ag back-reflection coating, is highly reflective and the sun image can be clearly observed when the photo was taken outdoor. (b) Measured (solid) and simulated (dash) spectral emissivity of the proposed LiF cooler in the IR range. Spectral emissivity requirement of the ideal cooler and the atmospheric transmittance are also provided. (c) Measured (solid) and simulated (dash) spectral absorptance of the LiF cooler within the solar irradiation spectrum between 0.3 and 2.5 μm. Inset is the schematic of the LiF

cooler. Spectral emissivity requirement of the ideal cooler is plotted as a black dash line, and the AM1.5 global tilted solar irradiation is provided as the background.

A 1-mm-thick bulk LiF crystal with a diameter of 40 mm is shown in Figure 1a (left in the figure), which is highly transparent in the visible wavelength range. The LiF sample (Shanghai Institute of Optics and Fine Mechanics, Chinese Academy of Sciences) was grown and crystallized from melted LiF powder at a high temperature, and then polished on both sides with a mean surface roughness less than 10 nm. As the emitter base, the LiF crystal possesses supreme optical properties, i.e., selective infrared emission with a high emittance exactly within the atmospheric window and an extremly low solar absorptance (see detailed optical properties of LiF in Part I of Supplemental Materials). For the convenience of demonstrating sub-ambient daytime cooling effect in experiment, a 200-nm-thick Ag film was deposited onto the backside of the LiF crystal through electron beam evaporation to achieve opaqueness through the whole spectrum. An ultraviolet-visible-near infrared spectrophotometer and a Fourier-transform infrared spectroscopy were employed to measure the directional, spectral transmittance $T_\lambda$ and reflectance $R_\lambda$ of the bulk LiF cooler at a wavelength range of 0.3-2.5 μm and 2.5-25 μm, respectively. The absorptivity $\alpha_\lambda$ and emissivity $\varepsilon_\lambda$ were obtained from $\alpha_\lambda = 1- R_\lambda$ and $\varepsilon_\lambda = \alpha_\lambda$ according to the Kirchhoff's Law. A specular surface was assumed because the LiF crystal surfaces were well polished with a mean surface roughness less than 10 nm, which could also be observed from the clear sun image shown in Figure 1a (right in the figure). We want to emphasize that a bulk LiF crystal is the main body to achieve sub-ambient daytime radiative cooling, which also guarantees the potential broad applications in optical systems as mentioned above, and the reason to add Ag back coating is for the convenience of experimental measurement in this work.

The properties of our proposed LiF cooler, as well as the ideal cooler, are compared in Figure 1b and c. For the infrared spectrum, the LiF cooler exhibits excellent selective emission exclusively in the wavelength range between 8 and 13 μm, which perfectly matches the atmospheric window with high atmospheric transmittance. The characteristic selective emission indicates great potential of the LiF cooler to be employed for sub-ambient radiative cooling effect. For the solar spectrum, the LiF cooler shows a pretty low absorptance beyond the wavelength of 0.4 μm, and the enhanced absorption below 0.4 μm is due to the intrinsic absorption of Ag coating. The total absorptance calculated in the solar spectrum is only 0.047 based on the AM1.5 global tilted data,[30] which is significant for the daytime cooling purpose. Based on the Ray Tracing method [31] and optical properties from Palik's data,[32] we also theoretically calculated the radiative properties of the LiF cooler, which are provided in Figure 1b and c as the dash lines. A perfect match between the results from the theoretical calculation and experimental measurement could be clearly observed.

For the comparison purpose, the ideal cooler is supposed to occupy a unity emissivity only in the wavelength range between 8 to 13 μm, and a zero absorptance in the solar spectrum. The working principle of the radiative cooler is schematically displayed in the inset of Figure 1c. Heat is dissipated to the cold space by emitting radiation through the atmospheric window in the wavelength range between 8 and 13 μm. Radiation at other wavelengths is considered undesired heat gain and is reflected by the cooler. The property of our proposed LiF cooler is close to that of the ideal cooler, especially the nearly ideal infrared selection, which is even advantageous to well-designed photonic structures. The simple construction and superior spectral selectivity of our proposed cooler provide it great potential in wide applications of sub-ambient daytime radiative cooling.

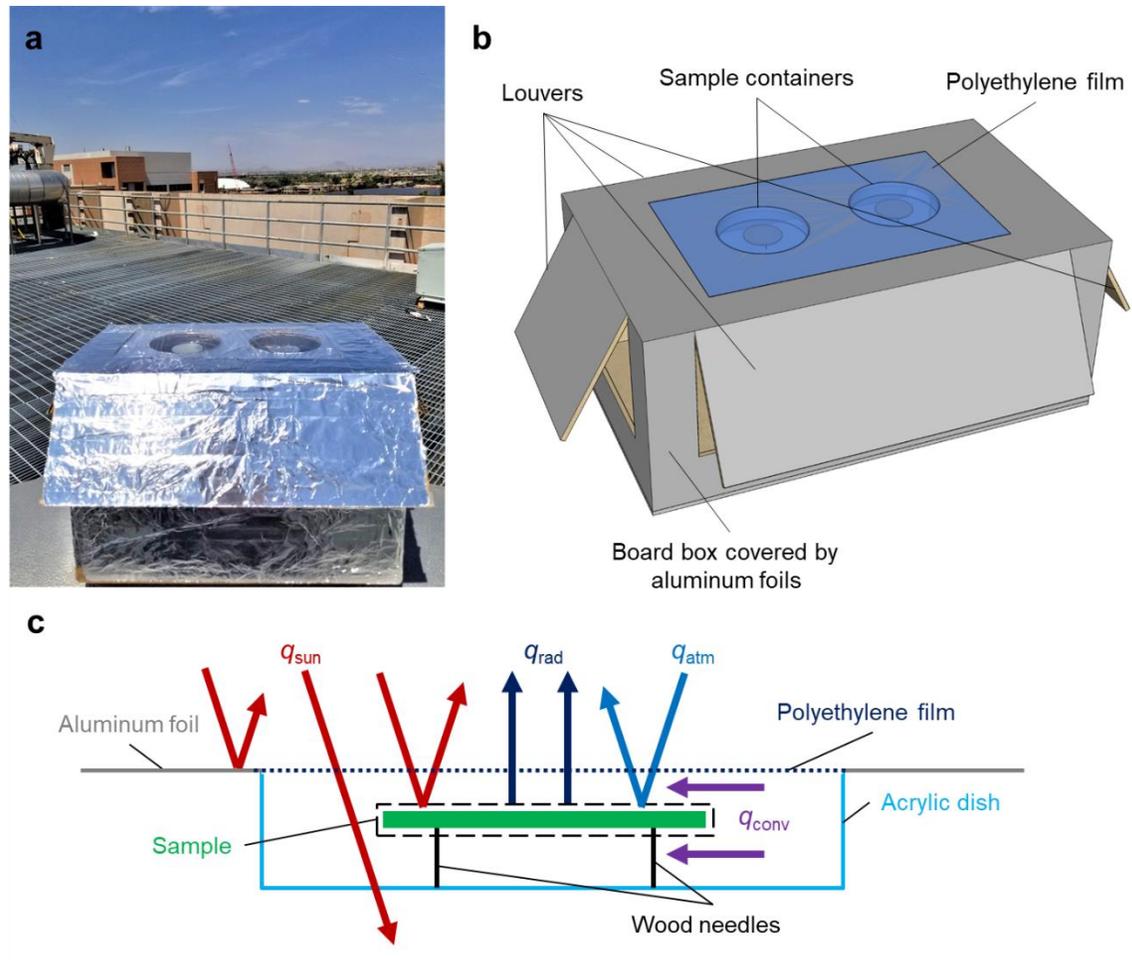

**Figure 2. Apparatus for outdoor test of radiative cooling performance.** (a) Photo of the apparatus on a rooftop in Tempe, Arizona. (b) Three-dimensional schematic of the apparatus. The apparatus was designed for providing full access to the sky and horizontal sunlight, as well as minimizing the parasitic heat gain to the sample through convection and conduction. (c) Cross-section schematic of the sample container and the heat transfer process related to the sample.

We then conducted an outdoor test to experimentally demonstrate the daytime cooling performance of the proposed cooler under direct sunlight. The apparatus was placed on a rooftop in Tempe, Arizona, and a photo of the experimental apparatus is shown in Figure 2a. The apparatus is schematically plotted in Figure 2b to show the detailed design. Two 4-inch sample containers on top of a board box allowed us to test two samples simultaneously. The sample containers were made of top-open clear acrylic dishes and were covered by a polyethylene film, which acted as a solar- and infrared-transparent wind shield. The containers were sealed to the underside of the top surface of the box, making the sides and bottom in full contact with the ambient air. The board box was wrapped by aluminum foils to reduce the solar heat gain, and louvers were created on each side of the box to allow free air flow and to shade direct sunshine. The idea of this apparatus was to minimize the parasitic heat gain of the sample through convection and conduction. The zoomed-in schematic of the sample container, as well as the related heat transfer process, is provided in Figure 2c. The sample was placed horizontally on three wood needles to minimize the heat conduction from the dish. With the coverage of the polyethylene film, the convectional heat transfer was limited to natural convection instead of forced convection around the sample. By using this apparatus, we were able to maintain the sample facing up with full access to the sky and exposed to strong horizontal solar illumination without shades. A pre-calibrated thermistor with an accuracy of 0.1°C was mounted at the center of the back surface of the sample, and the ambient temperature was measured by thermistors in the board box without sunshine and with free air flow. The temperatures were connected to a data logger (Omega OM-USB-TEMP), and the global solar irradiation was measured by a solar power meter (Amprobe SOLAR-100) placed horizontally near the apparatus.

The measured sample stagnation temperature could be derived under quasi-steady state condition, as shown in Figure 2c, through an energy balance equation of the sample

$$q_{rad}(T_s) - q_{sun}(G) - q_{atm}(T_a) - q_{conv}(T_s, T_a) = 0 \qquad (1)$$

In this equation, $q_{rad}$ is the radiative power emitted by the sample; $q_{sun}$ and $q_{atm}$ denote the received radiation from the sun and atmosphere, respectively; $q_{conv}$ is the heat gain from the surrounding air through convection. Each term is elucidated in Part II of Supplemental Materials. It can be seen that for a quasi-steady state, the sample stagnation temperature $T_s$ is highly dependent on the ambient temperature $T_a$, convective heat transfer and solar irradiance $G$.

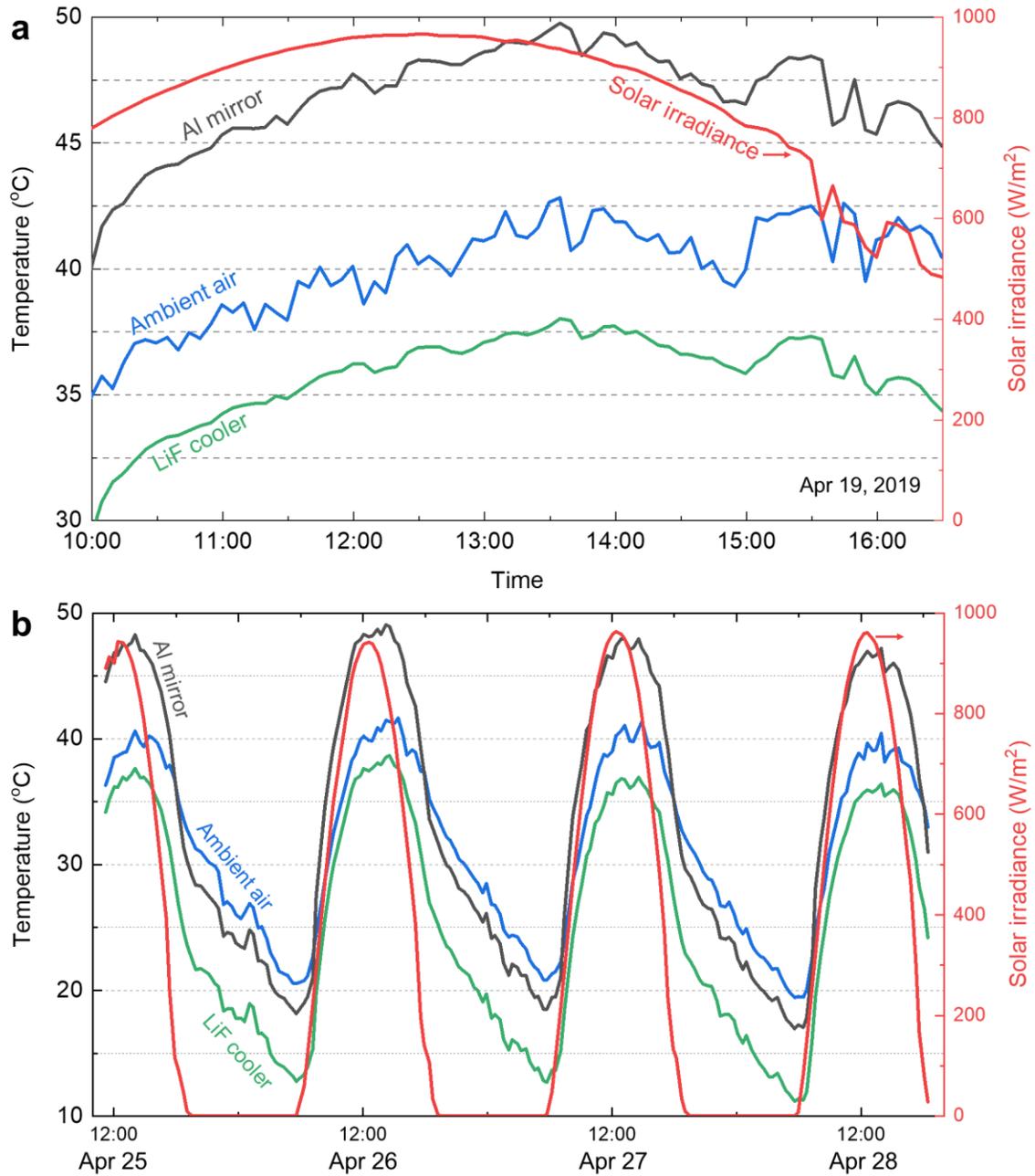

**Figure 3. Results of the outdoor test on a rooftop** in Tempe, Arizona. Measured stagnation temperatures of the proposed LiF cooler, Al mirror and ambient air, and solar irradiance on a horizontal surface: (a) on a sunny day (April 19, 2019), and (b) for four consecutive days (April 25-28, 2019).

The LiF cooler placed in the container is of 40 mm in diameter and has a total thickness of 1

mm. In addition to our LiF cooler, an aluminum mirror coated on a 1-mm-thick glass, was tested simultaneously in the other container as a reference. The measured temperatures and solar irradiance are plotted in Figure 3, where results on different time periods are provided. On the sunny days in April of Arizona, the ambient temperature can reach above 40 °C, and the peak solar irradiance is over 960 W/m². In spite of a low solar absorptance of 0.074 the Al mirror, as shown in Figure 3a, is much hotter than the ambient air during the daytime. The highest temperature of the Al mirror reaches nearly 50°C, which is approximately 8°C above the ambient temperature. This is mainly due to the low infrared emissivity of 0.012 from the Al mirror (assuming the mirror temperature as 45°C). On the other hand, due to the nearly ideal infrared emission exclusively within the atmospheric window, our proposed LiF cooler is able to dissipate heat directly to the cold space by passive thermal emission in addition to its low solar absorptance of 0.047. As a result, its stagnation temperature is always lower than the ambient air temperature during the daytime under direct sunlight. The temperature drop could be up to 5°C below the ambient during the hottest period of the day from 10:00 am to 4:00 pm with a high solar irradiance exceeding 900 W/m² around noon time (See Fig. S2 for the theoretical calculation that verifies the experimental results).

Radiative cooling measurement for four consecutive days was also performed to show the performance in a longer time period, as shown in Figure 3b. The LiF cooler demonstrates stable cooling performance constantly below the ambient air in the daytime (i.e. 7 am to 6 pm) with peak solar irradiance above 960 W/m², and the temperature drop can be close to 10°C during the nighttime. As a comparison, the Al mirror is always hotter than the ambient air during the daytime, while it could gain a stagnation temperature below the ambient at night by1~2°C.

In addition to the sub-ambient temperature drop, the cooling power $q_{cool}$ is another major figure of merit for evaluating the radiative cooling performance, which is given by

$$q_{cool}(T_s) = q_{rad}(T_s) - q_{sun}(G) - q_{atm}(T_a) - q_{conv}(T_s, T_a) \qquad (2)$$

To directly measure the net cooling power $q_{cool}(T_s = T_a)$, as depicted in the inset of Figure 4a, a thin-film resistive heater was attached to the underside of the LiF cooler through a 1.5-mm-thick copper plate, and electric power $q_{heater}$ was supplied to the heater for generating heat through a direct-current power source (Keithley 2230-30-1), which is the cooling load at steady state, i.e., $q_{heater} = q_{cool}$. We measured the net cooling power produced by the proposed LiF cooler in the daytime and nighttime. For the cooling power measurement, a steady state was reached first with $q_{heater} = 0$. We then started with a small amount of input power $q_{heater}$ and increased the power in several steps. Each power was maintained constantly for 10 minutes or longer, and the quantities and time of $q_{heater}$ are indicated in Figure 4. By tracking the emitter and ambient temperature, we were able to recognize the time when a quasi-steady state was reached, and $q_{heater}$ at this condition where the emitter temperature equals to the ambient temperature represents the net cooling power $q_{cool}$ ($T_s = T_a$) produced by the proposed cooler.

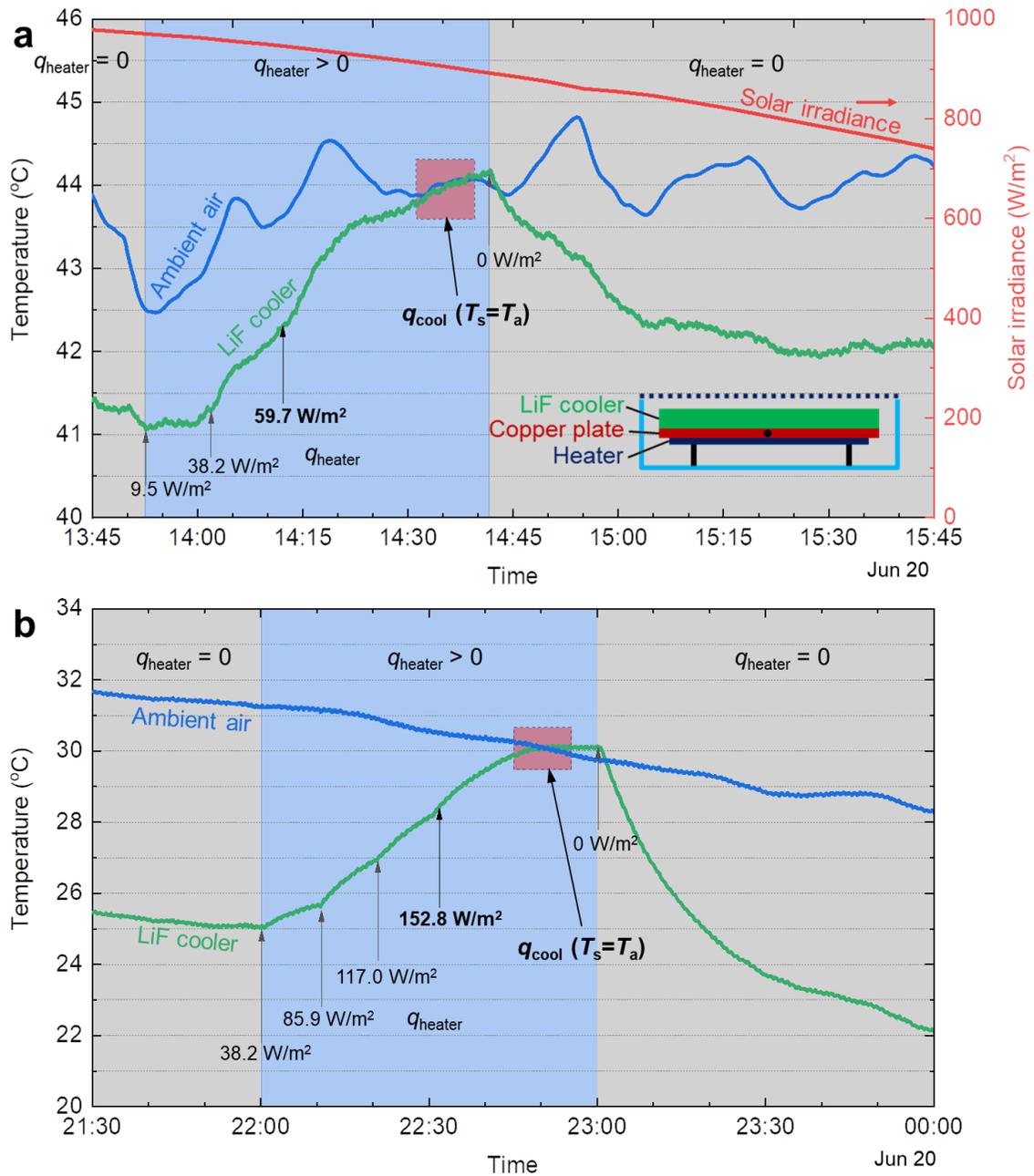

**Figure 4. Net cooling power measurement of the LiF cooler during** (a) the daytime and (b) nighttime. A schematic of the cooling power measurement setup is provided in the inset of (a). During the measurement, a certain amount of electric power was supplied to the LiF cooler by an electric heater through a copper plate. Each power was maintained for 10 minutes or longer. The input electric power at the condition where the cooler temperature equals to the ambient temperature, represents the net cooling power produced by the cooler.

The experimental results displayed in Figure 4a represent the cooling power measurement in the afternoon. In the beginning, the cooler temperature was approximately 2.5°C below the ambient temperature with $q_{heater} = 0$. With an initial $q_{heater} = 9.5$ W/m$^2$, the cooler temperature raised slightly and kept approaching the ambient temperature as $q_{heater}$ increased. During the period when the cooler temperature equaled to the ambient one, we obtained a net cooling power of $q_{cool} = 59.7$ W/m$^2$ under a solar irradiance of 900 W/m$^2$ around 2:35 pm. A similar measurement was taken at night, as shown in Figure 4b, and a net cooling power of 152.8 W/m$^2$ was reached around 10:50 pm.

The proposed cooler, which is based on the bulk LiF material, displays decent cooling performance for both sub-ambient temperature drop and net cooling power. It should be pointed out that the temperature drop and cooling power are highly dependent on the local meteorological conditions such as the ambient temperature, solar irradiance, and parasitic heat loss. The effects of sky condition (e.g., clear, cloudy, etc.), humidity, and wind speed have been investigated and these could strongly affect the cooling performance.[33,34] Our outdoor experimental data have shown that the cooling performance varied on different days even for the same apparatus and the same samples due to different weather conditions (see Part III of Supplemental Materials). Therefore, neglecting the variations of experimental conditions in different studies, the spectral property of the cooler should be the core factor needing to be considered in terms of achieving better radiative cooling performance. Just from the aspect of the spectral property of a radiative cooler, the comparison of the sub-ambient cooling performance between our proposed LiF cooler and previous well-designed structures is elucidated in detail in Part IV of Supplemental Materials.

Sub-ambient daytime radiative cooling effects have been demonstrated through radiators with low solar absorption and high infrared emission. Although selective emitters are advantageous to broadband ones in terms of the potential to achieve an ultra-low stagnation temperature, the current designs are mainly based on photonic structures, whose applications are restricted by their high fabrication cost and complicated manufacturing process. In this work, we proposed a cooler consisting of a single layer of bulk LiF crystal and Ag back reflector, which gives a solar absorptance of 4.7% and nearly ideal infrared selectivity for sub-ambient daytime cooling purpose. The conducted outdoor test demonstrated a prominent sub-ambient cooling effect with a temperature drop of 5 °C below the ambient and a net cooling power of 60 W/m$^2$ under horizontal solar illumination exceeding 900 W/m$^2$. Furthermore, as a bulk material with the highest ultraviolet-visible-near infrared transmittance, LiF has been widely employed as the optical windows and mirrors in various applications. Compared to other optical windows or mirrors, the bulk LiF crystal occupies excellent achromatic and polychromatic properties, and its small refractive index provides it another advantage in terms of not requiring anti-reflection coatings. The demonstrated radiative cooling effect based on the bulk LiF material would open an innovative area to radiatively cool optical systems.

# Supplemental Materials

# Bulk Material Based Selective Infrared Emitter for Sub-Ambient Daytime Radiative Cooling

Yue Yang,[1,#,*] Linshuang Long,[2,#] Sheng Meng,[1] Nicholas Denisuk,[2] Liping Wang,[2,*] and Yonggang Zhu[1,3]

[1] School of Mechanical Engineering and Automation,
Harbin Institute of Technology, Shenzhen, 518055, PR China

[2] School for Engineering of Matter, Transport & Energy,
Arizona State University, Tempe, AZ 85287, USA

[3] School of Science, RMIT University, Melbourne, VIC 3001, Australia

[#] These authors contributed equally to this work.

[*]Corresponding author. Email: yangyue2017@hit.edu.cn (Y.Y.); liping.wang@asu.edu (L.W.)

## Part I. Optical Properties of Lithium Fluoride

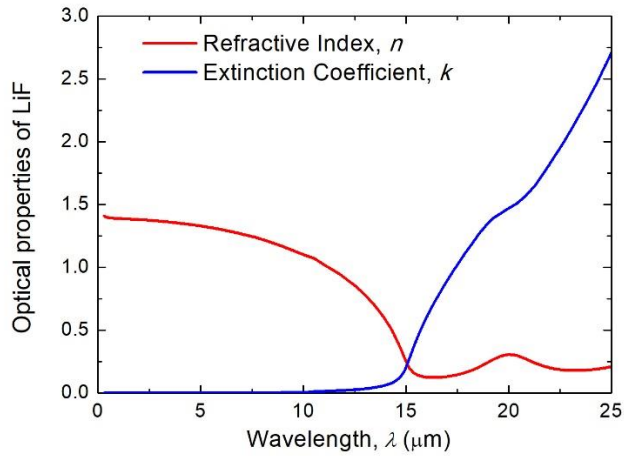

Figure S1. Optical properties, including the refractive index ($n$) and the extinction coefficient ($k$), of lithium fluoride (LiF) from solar to infrared spectrum.

The reason why we chose bulk LiF as the radiative cooler could be found in Figure S1, which plots the refractive index ($n$) and extinction coefficient ($k$) along the whole spectrum from solar to mid-infrared. These properties were taken from Palik's handbook.[1] As observed in Figure S1, in solar and near-infrared spectrum, $k$ is approaching zero, which indicates a high transmittance of the bulk LiF. With the help of Ag back coating, the bulk LiF emitter could obtain a high reflectance in solar and near-infrared spectrum. At wavelengths beyond 20 μm, $k$ is much greater than $n$, which implies that bulk LiF behaves like a metal with a high reflectance. However, for the mid-infrared region containing the atmospheric window, $k$ starts to increase and becomes comparable with $n$, which may cause a high emittance as desired.

## Part II. Theoretical Model for Heat Transfer Analysis

To better understand and validate the measurement from the outdoor tests, a theoretical model of heat transfer analysis has been established. The related heat transfer processes have been depicted in Figure 2c. Under quasi-steady state condition, the energy balance of

the sample yields

$$q_{rad}(T_s) - q_{sun}(G) - q_{atm}(T_a) - q_{conv}(T_s,T_a) = 0 \tag{S1}$$

In this equation, $q_{rad}$ is the radiation power emitted by the sample; $q_{sun}$ and $q_{atm}$ denote the received radiation from the sun and atmosphere, respectively; $q_{conv}$ is the heat gain from the surrounding air through convection.

The emitted power per unit area in Eq. (S1) is given by

$$q_{rad}(T_s) = \int_0^\infty \int_0^{2\pi} \int_0^{\pi/2} I_{BB}(\lambda,T_s)\varepsilon_{\lambda,\theta}(\lambda,\theta,\phi)\cos\theta\sin\theta\,d\theta\,d\phi\,d\lambda \tag{S2}$$

where $\varepsilon_{\lambda,\theta}(\lambda,\theta,\phi)$ is the spectral, directional emittance of the sample, and $I_{BB}(\lambda, T)$ is the spectral radiance of a blackbody at temperature $T$ according to the Planck distribution. The wavelength- and direction-dependent properties are specified in terms of wavelength $\lambda$, the zenith angle $\theta$, and azimuthal angle $\phi$.

The received radiative powers per unit area are

$$q_{sun} = G\alpha_{solar} \tag{S3}$$

$$q_{atm}(T_a) = \int_0^\infty \int_0^{2\pi} \int_0^{\pi/2} I_{BB}(\lambda,T_a)\alpha_{\lambda,\theta}(\lambda,\theta,\phi)\varepsilon_{atm}(\lambda,\theta)\cos\theta\sin\theta\,d\theta\,d\phi\,d\lambda \tag{S4}$$

where $G$ is the measured solar irradiance, $\alpha_{solar}$ is the integrated absorptance of the sample based on the AM1.5 global title data,[2] and $\varepsilon_{atm}(\lambda,\theta)$ is the spectral, directional emittance of the atmosphere. $\varepsilon_{atm}(\lambda,\theta)$ can be estimated through $\varepsilon_{atm}(\lambda,\theta) = 1 - t(\lambda)^{1/\cos\theta}$ with the atmospheric transmittance in the zenith direction $t(\lambda)$.[3,4]

The convectional power per unit area between the sample temperature $T_s$ and ambient air temperature $T_a$ is

$$q_{conv}(T_s,T_a) = 2h_c(T_a - T_s) \tag{S5}$$

where $h_c$ is the convection heat transfer coefficient, and the factor of 2 means the heat convection occurs at both top and bottom sides of the sample as the sample is suspended. Here we employed the value of 6.9 W/m² K for each side, which was a typical value commonly used in similar setups.[5]

Among these equations, several assumptions have been made to further simplify the analysis:

1. Other radiative heat transfer modes such as that between the sample and the acrylic dish, and radiation from sample bottom surface are neglected due to the relatively small temperature difference.
2. The conduction is neglected as the sample is well insulated with wood needles.
3. The sample temperature is uniform due to the thin thickness of the sample.
4. The sample surface is specular as the surface is well polished so that the directional dependency of the radiative properties can be ignored.
5. A constant transmittance of 0.9 was considered for the PE film cover in the theoretical model through a Ray Tracing method.[6]

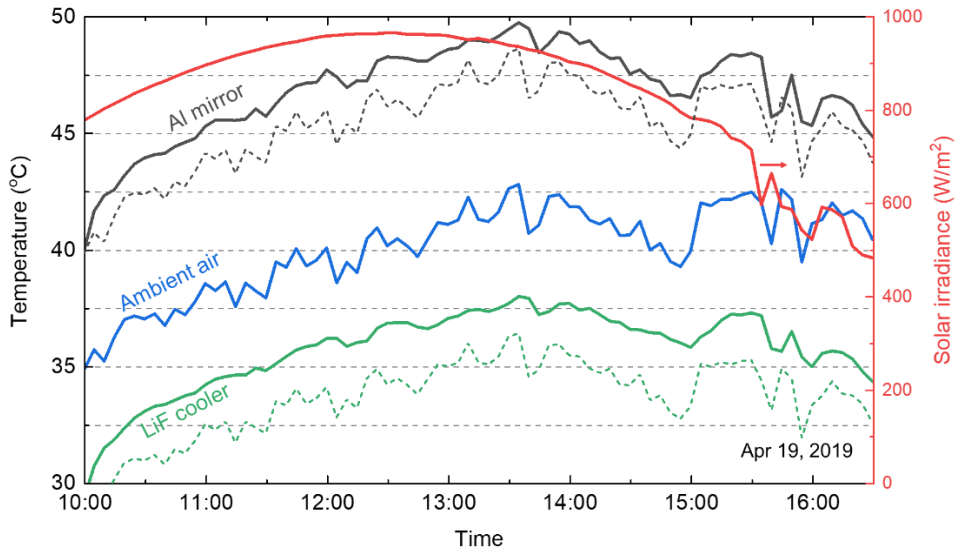

Figure S2. Comparison of the temperatures between the experimental measurement and theoretical calculation. Solid lines: measurement from the daytime outdoor test. Dash lines: calculation from the theoretical heat transfer model.

With the measured solar irradiation $G$, ambient air temperature $T_a$, and radiative properties, the stagnation temperatures for both Al mirror and LiF cooler at different time can be calculated based on the theoretical model and are presented in Figure S2 as dash lines, which agree with experimental data quite well. The assumptions mentioned above for the theoretical analysis are responsible for the temperature difference of 1~2 ℃ between the experimental measurement and theoretical calculation for both Al mirror and LiF cooler samples. In the theoretical heat transfer model, we only considered the radiative heat transfer between the sample and the ambient air, while the radiative heat gain from the surroundings were neglected, which could give a higher stagnation temperature in experiment as shown in Fig. S2. In addition, it should be noticed that although a typical convection coefficient value of 6.9 W/m² K was employed in the theoretical model, it may vary a lot in a specific experimental setup and with a different ambient temperature. Furthermore, the solar absorptance applied in the theoretical model was based on the AM1.5 global tilted data, but the solar irradiance measured in Arizona was different. Therefore, with considerations of all the variations mentioned above, a good measurement accuracy could be confirmed even with a 1~2 ℃ temperature difference.

Based on the energy balance equation of Eq. (S1), the radiative cooling power of the sample is given by

$$q_{cool}(T_s) = q_{rad}(T_s) - q_{sun}(G) - q_{atm}(T_a) - q_{conv}(T_s, T_a) \tag{S6}$$

which is the outgoing radiative power subtracting the incoming power. It can be seen that, in order to achieve a sub-ambient $T_s$, the net cooling power of the sample should be positive with $T_s = T_a$, which means the radiative power emitted by the sample is larger than the heat gain. To measure the net cooling power $q_{cool}(T_s = T_a)$, electric power $q_{heater}$ is supplied to the sample through the heater to maintain $T_s = T_a$. When a new steady state is reached, the energy balance equation becomes

$$q_{rad}(T_s = T_a) - q_{sun}(G) - q_{atm}(T_a) - q_{heater} = 0 \tag{S7}$$

where the convectional item $q_{conv}$ is eliminated as $T_s = T_a$. Comparing Eq.(S6) and (S7), we have $q_{cool}(T_s = T_a) = q_{heater}$.

**Part III. Additional Experimental Results in Outdoor Tests**

As mentioned in the manuscript, the merits for evaluating cooling performance, the sub-ambient temperature drop and net cooling power, are highly dependent on the weather conditions, e.g., solar irradiance, sky conditions, ambient temperature, humidity, and wind speed. Here we provide some experimental data we measured on different days to show that the cooling performance varies on different days even for the same apparatus and same samples due to different weather conditions.

Figure S3 shows the temperature measurement, as well as the photo of the sky condition, on different days. We can see that when the sky is cloudy, the temperature drop of the cooler compared to the ambient air is small, and sometimes there is no temperature drop at all. Generally, a clear sky is desired to demonstrate the cooling performance.

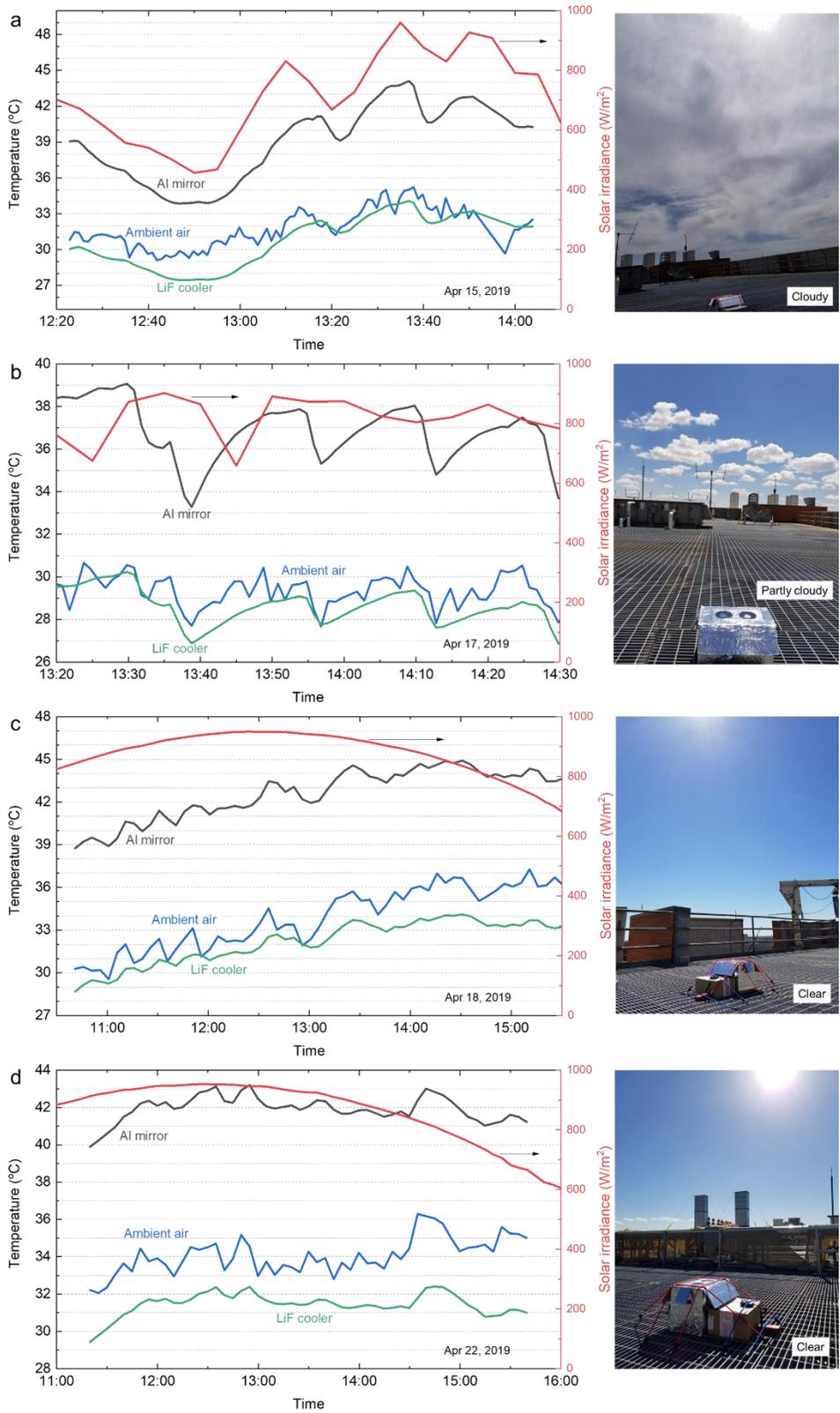

Figure S3. Outdoor tests on different days. The temperature and solar irradiance measurements are shown in the left column. The photos of the sky conditions are provided

in the right column.

Figure S4 displays the cooling power measurement on another day different from the one shown in Figure 4 of the manuscript. The net cooling power is around 50 W/m$^2$, which is 10 W/m$^2$ less than the measurement discussed in the manuscript due to varied weather conditions on different days. A photo of the cooling power measurement is provided in Figure S5.

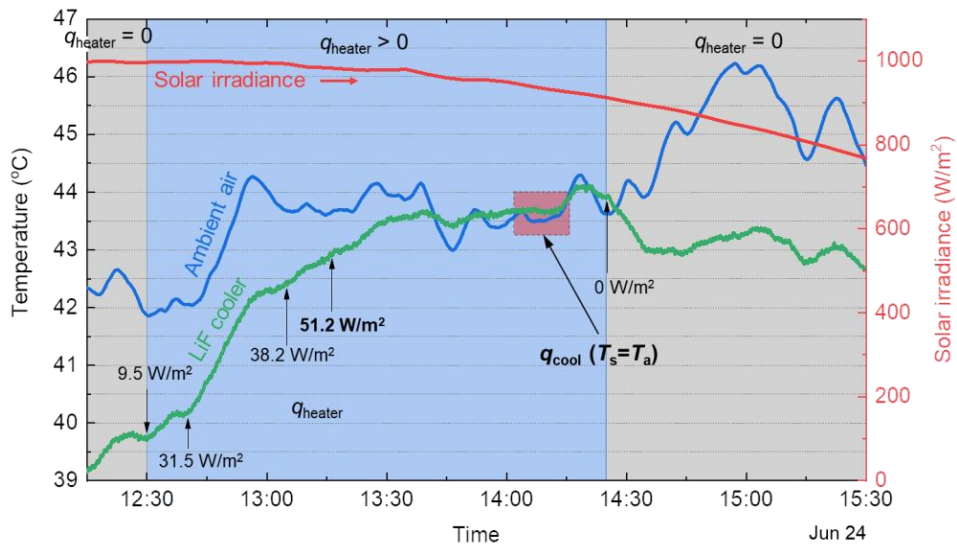

Figure S4. Cooling power measurement on another day different from the one reported in the manuscript.

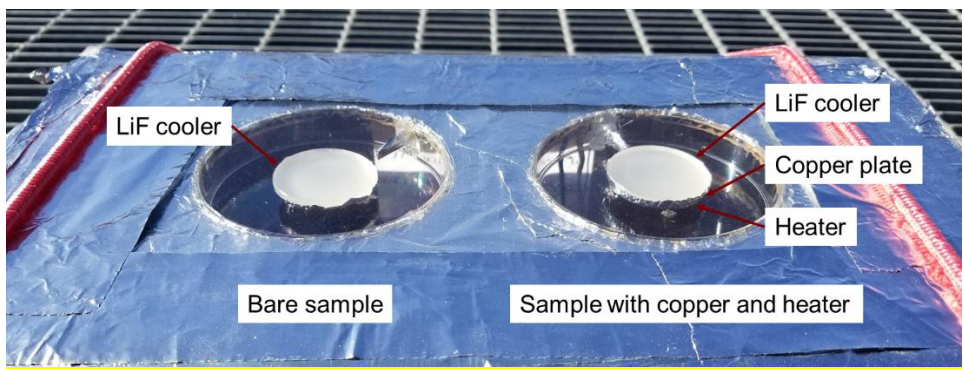

Figure S5. Photo of the apparatus for cooling power measurement.

Considering that the cooling perfomance varies strongly upon the weather condition even for the same setup and structure, we believe that we could also observe another temperature drop or cooling power larger than the ones reported in the manucript. Therefore, it is not fair to judge the performance of a radiative cooler just based on the experimental measurement from "selected" ourdoor tests. Neglecting the variations of experimental conditions in different studies, the spectral property of the cooler should be the core factor needing to be considered in terms of achieving better radiative cooling performance. Just from the aspect of the spectral property of a radiative cooler, the comparison of the sub-ambient cooling performance between our proposed LiF cooler and previous well-designed structures is detailly elucidated in the next part.

**Part IV. Comparison Between Our Proposed Emitter with Ideal Emitters, and Other Designs**

As we mentioned in the manuscript, the reported best daytime cooling performance up to date from outdoor tests conducted under direct sunlight is within a range of 4~8 °C below the ambient temperature and 40~130 W/m$^2$ net cooling power.[5,7-10] However, considering the variations of environmental conditions and experimental setups, which could significantly affect the cooling performance,[11] it is hard to judge the advantages of any material over others just from stagnation temperature or cooling power measurement. Therefore, to obtain a reasonable comparison between different designs from the material aspect, as shown in Figure S6, we theoretically calculated the stagnation temperature and net cooling power for all designs in the same environmental conditions. The ambient temperature was set as 300 K, and a AM1.5 solar irradiation was assumed to directly shine on the emitters, including our proposed emitter, the ideal emitters and other ones,[5,7-10] which have demonstrated sub-ambient daytime cooling effect in outdoor tests. The ideal selective emitter indicates an emitter with unity emittance only at the atmospheric window from 8 to 13 μm, while the ideal broadband emitter represents the one with unity emittance in the whole mid-infrared region from 4 to 25 μm. The heat transfer model used for theoretical calculation was the same with that in the previous part. Note that in Figure S6(b), the parasitic heat loss is eliminated with the convective heat transfer coefficient equal to zero.

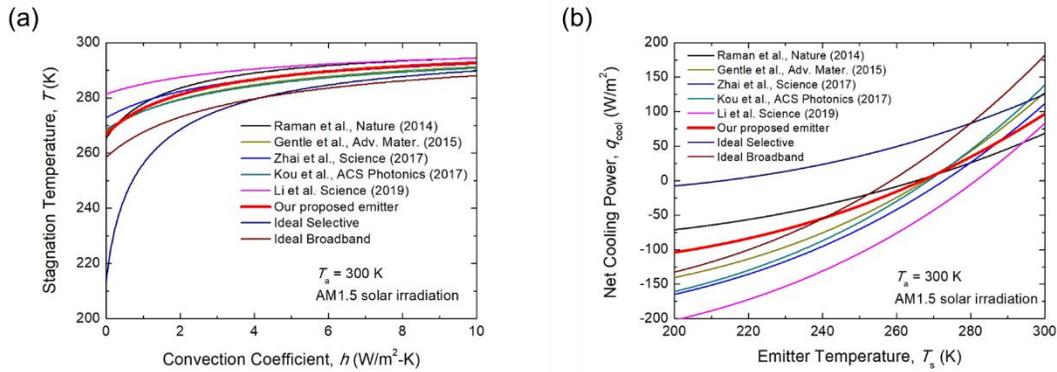

Figure S6. Comparison of (a) achieved stagnation temperature and (b) net cooling power between our proposed emitter and ideal emitters, and other works as well, including Raman et al. Nature (2014) [5]; Gentle et al., Adv. Mater. (2015) [7]; Zhai et al., Science (2017) [9]; Kou et al., ACS Photonics (2017) [8]; Li et al., Science (2019) [10]. Note that all these works compared above have demonstrated sub-ambient daytime cooling effect in outdoor tests.

We can clearly observe that in Figure S6, ideal selective emitter occupies unrivalled advantage in achieving ultra-low stagnation temperature, which could be below the ambient one as much as 85 °C with eliminated convective loss, far beyond the limit of ideal broaband one with 40 °C temperature drop. Our proposed emitter could also potentially obtain a stagnation temperature over 30 °C below the ambient one, which is among the best ones compared with other designs shown in Figure S6(a). When the emitter temperature is set the same with the ambient one, ideal broadband emitter provides the highest net cooling power around 175 W/m$^2$, while our proposed emitter could also give a cooling power over 100 W/m$^2$. Although the relatively high solar absorptance of our proposed emitter (around 5%) makes it disadvantegeous in daytime radiative cooling under direct sunlight, it still demonstrates a close or even better cooling performance compared to the up-to-date top-level radiative emitters.

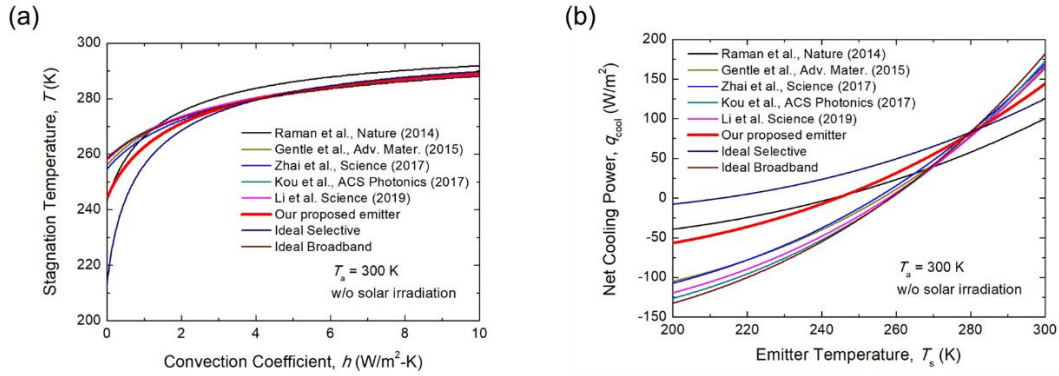

Figure S7. Comparison of (a) achieved stagnation temperature and (b) net cooling power between our proposed emitter and ideal emitters, and other works as well. Note that, different from Figure S6, the solar irradiation is eliminated here, corresponding to the applications of space cooling, nighttime cooling, or daytime cooling with direct sunlight blocked.

In addition, the nearly ideal selective infrared emission and simple fabrication process of our proposed emitter makes it superior to any existing design in applications with solar irradiation eliminated, like space cooling, nighttime cooling, or daytime cooling with direct sunlight blocked. As observed in Figure S7 where solar absorption is eliminated, our proposed emitter has unrivalled advantage in radiative cooling purpose compared to other designs. It could gain the lowest stagnation temperature over 55 °C below the ambient one. Although the photonic crystal based emitter proposed in [5] could potentially achieve a similar temperature drop with eliminated convective loss, the net cooling power provided, i.e. 100 W/m$^2$, is much less than that given by our proposed emitter, i.e. 150 W/m$^2$. Moreover, the bulk layer with metal coating structure of our proposed emitter holds another advantage of simple fabrication process and scalable manufacturing ability compared to photonic structure based ones.